\documentstyle[aps,preprint,tighten]{revtex}
\newcommand{\select}[2]{#1}

\begin{document}

\draft
\preprint{KAIST-TH 99/07, hep-th/9909053}
\title{Inflation and Gauge Hierarchy in Randall-Sundrum Compactification}
\author{Hang Bae Kim$^{a\dagger}$ and Hyung Do Kim$^{b\ddagger}$}
\address{$^a$Department of Physics and Institute of Basic Science\\
Sungkyunkwan University, Suwon 440-746, Korea\\
$^b$Department of Physics, Korea Advanced Institute of Science and Technology\\
Taejon 305-701, Korea\\
$^\dagger${\tt hbkim@newton.skku.ac.kr},
$^\ddagger${\tt hdkim@muon.kaist.ac.kr}}
\maketitle
\begin{abstract}
We obtain the general inflationary solutions for the slab of five-dimensional
AdS spacetime where the fifth dimension is an orbifold $S^1/Z_2$ and
two three-branes reside at its boundaries,
of which the Randall-Sundrum model corresponds to the static limit.
The investigation of the general solutions and their static limit
reveals that
the RS model recasts both the cosmological constant problem and the gauge
hierarchy problem into the balancing problem of the bulk and the brane
cosmological constants.
\end{abstract}

\select{}{\begin{multicols}{2}}


The huge gap between the electroweak scale $M_{\rm W}$ and the Planck scale 
$M_{\rm P}$, $M_{\rm P}/M_{\rm W}\sim10^{16}$, has been a long standing puzzle
in unifying the standard model and the gravity.
Recently the extra dimensional models addressing it have drawn much attention
\cite{ADD,RS1}.
Introducing the extra dimensions has a long history from Kaluza-Klein theory
to string theory.  What is new is that the
ordinary matter is confined to our four-dimensional brane while the gravity
propagates in the whole spacetime.
One motivation for such models is the heterotic M theory,
whose field theory limit is the 11-dimensional supergravity compactified
on $S_1/Z_2$ with supersymmetric Yang-Mills fields living on two boundaries
\cite{HW}.

The large extra dimension model \cite{ADD} brings the fundamental
gravitational scale around the weak scale to solve the gauge hierarchy problem
and reduce the strength of the 4-dimensional gravity by having the large extra
dimensions, while avoiding conflict with experiments by confining the standard
model fields to a 3-brane in the extra dimensions.
This model translates the gauge hierarchy to the hierarchy between the
fundamental scale and the size of the extra dimensions.

More recently, Randall and Sundrum (RS) proposed a five-dimensional model
with nonfactorizable geometry supported by the negative bulk cosmological
constant and the oppositely signed boundary 3-brane cosmological constants.
In this model, the gauge hierarchy problem can be explained by 
the exponential warp factor even for the small extra dimension.
The model is quite interesting and has drawn much attention
because it might be realizable
in supergravity and superstring compactifications
\cite{HW,CS,Kehagias,Verlinde}.
However, its geometry is based on the very specific relation between
the bulk and the brane cosmological constants.
Then the question arises how precisely this relation should hold to preserve
the necessary geometry.
In this Letter, we try to answer this question
by finding the cosmological inflationary solutions
with general sets of the bulk and the brane cosmological constants
and comparing them with the static solution given by RS.
The cosmological aspect of the extra dimensional models has been discussed
by many authors \cite{BDL,CGKT,CGS}.
Especially the inflationary solutions were obtained
for the flat bulk geometry \cite{LOW},
and for the AdS bulk geometry \cite{Nihei,Kaloper},
where the condition is imposed among parameters
such that the extra dimension does not inflate.
Here we obtain the general inflationary solutions for the AdS bulk
geometry and focus on the connection to the gauge hierarchy problem.


We consider the five-dimensional spacetime with coordinates $(\tau,x^i,y)$
where $\tau$ and $x^i, i=1,2,3$ denote the usual four-dimensional spacetime
and $y$ is coordinate of the fifth dimension,
which is an orbifold $S^1/Z_2$ where the $Z_2$ action identifies $y$ and $-y$.
We choose the range of $y$ to be from $-1/2$ to $1/2$.
We consider two 3-branes extending in the usual 4-dimensional spacetime
reside at two orbifold fixed points $y=0$ and $y=1/2$,
so that they form the boundaries of the five-dimensional spacetime.
This five-dimensional model is described by the action
\begin{eqnarray}
S &=& \int_{M}d^5x\sqrt{-g}\left[\frac{M^3}{2}R-\Lambda_b\right]
\select{}{\nonumber\\&&}
+ \sum_{i=1,2}\int_{\partial M^{(i)}}d^4x
    \sqrt{-g^{(i)}}\left[{\cal L}_i-\Lambda_i\right],
\end{eqnarray}
where $M$ is the fundamental gravitational scale of the model,
$\Lambda_b$ and $\Lambda_i$ are the bulk and the brane cosmological constants,
${\cal L}_i$ are the Lagrangians for the fields confined in the branes.
Since we are interested in the cosmological solution, we assume that
the three-dimensional spatial section is homogeneous and isotropic.
Further we consider it is flat for simplicity.
The most general metric satisfying this can be written as
\begin{equation}
\label{eq:metric}
ds^2 = -n^2(\tau,y)d\tau^2 + a^2(\tau,y)\delta_{ij}dx^idx^j + b^2(\tau,y)dy^2
\end{equation}
For the above action and metric, we obtain the following Einstein equations
corresponding to (00), (ii), (55), (05) components respectively:
\begin{eqnarray}
\label{Ee:00}
&&\frac{3}{n^2} \frac{\dot a}{a} \left(\frac{\dot a}{a}+\frac{\dot b}{b}\right)
-\frac{3}{b^2}
\left[\frac{a''}{a}+\frac{a'}{a}\left(\frac{a'}{a}-\frac{b'}{b}\right)\right]
\nonumber\\ && \hspace{0mm}
= M^{-3}\left[\Lambda_b
+\frac{\delta(y)}{b}\left(\Lambda_1+\rho_1\right)
+\frac{\delta(y-\frac12)}{b}\left(\Lambda_2+\rho_2\right)\right],
\\
\label{Ee:ii}
&&\frac{1}{n^2} \left[2\frac{\ddot a}{a}+\frac{\ddot b}{b}
-\frac{\dot a}{a}\left(2\frac{\dot n}{n}-\frac{\dot a}{a}\right)
-\frac{\dot b}{b}\left(\frac{\dot n}{n}-2\frac{\dot a}{a}\right)\right]
\nonumber\\&&
-\frac{1}{b^2} \left[\frac{n''}{n}+2\frac{a''}{a}
+\frac{a'}{a}\left(2\frac{n'}{n}+\frac{a'}{a}\right)
-\frac{b'}{b}\left(\frac{n'}{n}+2\frac{a'}{a}\right)\right]
\nonumber\\ && \hspace{0mm}
= M^{-3}\left[\Lambda_b
+\frac{\delta(y)}{b}\left(\Lambda_1-p_1\right)
+\frac{\delta(y-\frac12)}{b}\left(\Lambda_2-p_2\right)\right],
\\
\label{Ee:55}
&&\frac{3}{n^2} \left[\frac{\ddot a}{a}
-\frac{\dot a}{a}\left(\frac{\dot n}{n}-\frac{\dot a}{a}\right)\right]
-\frac{3}{b^2}\frac{a'}{a}\left(\frac{n'}{n}+\frac{a'}{a}\right)
= M^{-3}\Lambda_b,
\\
\label{Ee:05}
&&3\left(\frac{\dot a}{a}\frac{n'}{n}+\frac{\dot b}{b}\frac{a'}{a}
-\frac{\dot a'}{a}\right)
= 0,
\end{eqnarray}
where the dot and the prime represent the derivatives with respect to $\tau$
and $y$, respectively.
The equations with bulk and boundary sources are equivalent to
the equations with bulk sources and proper boundary conditions.
To give non-singular geometry, $n$, $a$ and $b$ must be continuous
along the extra dimension.  But (\ref{Ee:00}) and (\ref{Ee:ii}) imply that
$n'$ and $a'$ are discontinuous at $y=0,\pm\frac12$ so that $n''$ and $a''$
have delta function singularities there.
Applying $\int_{0^-}^{0^+}dy$ and $\int_{\frac12^-}^{\frac12^+=-\frac12^+}dy$
to (\ref{Ee:00}) and (\ref{Ee:ii}), we obtain the boundary conditions
\begin{eqnarray}
&&\left.\frac{n'}{n}\right|_{0^-}^{0^+}
	= -\frac{b(\tau,0)}{3M^3}(\Lambda_1-2\rho_1-3p_1),
\select{\quad}{\nonumber\\&&}
\left.\frac{a'}{a}\right|_{0^-}^{0^+}
	= -\frac{b(\tau,0)}{3M^3}(\Lambda_1+\rho_1)
\nonumber\\
&&\left.\frac{n'}{n}\right|_{-\frac12}^{+\frac12}
	= +\frac{b(\tau,\frac12)}{3M^3}(\Lambda_2-2\rho_2-3p_2),
\select{\quad}{\nonumber\\&&}
\left.\frac{a'}{a}\right|_{-\frac12}^{+\frac12}
	= +\frac{b(\tau,\frac12)}{3M^3}(\Lambda_2+\rho_2).
\label{BC}
\end{eqnarray}
Since $b''$ does not appear in the equations,
no boundary condition is imposed on $b'$.

The equations (\ref{Ee:00})--(\ref{Ee:05}) and the boundary conditions
(\ref{BC}) constitute a starting point of cosmology of the five-dimensional
model considered in this paper.
It's difficult to solve the whole bulk equations with generic sources,
but at the brane boundaries, the (55) and (05) equations give the
Friedman-like equation
\begin{equation}
\label{Friedman-equation}
\left(\frac{\dot a}{a}\right)^2 =
 \left(\frac{\Lambda_b}{6M^3}+\frac{\Lambda_i^2}{36M^6}\right)
+\frac{\Lambda_i}{18M^6}\rho_i
+\frac{1}{36M^6}\rho_i^2.
\end{equation}
The implications of this equation are quite interesting \cite{BDL,CGKT,CGS},
but we will not pursue them here.

In this paper, we consider the cosmological constant dominated cases,
and neglect the matter and radiation energy densities of the branes.
Then the above equations allow a static solution,
the so-called RS solution \cite{RS1},
when the bulk cosmological constant is negative ($\Lambda_b<0$) and related
to the brane cosmological constants by
\begin{equation}
\label{RS-condition}
k = k_1 = -k_2,
\end{equation}
where $k=(-\Lambda_b/6M^3)^{1/2}$ and $k_i=\Lambda_i/6M^3$.
We take the brane with the positive cosmological constant to be at $y=0$.
The metric of the static solution is given by
\begin{equation}
\label{RS-metric}
ds^2 = e^{-2kb_0|y|}\eta_{\mu\nu}dx^\mu dx^\nu + b_0^2dy^2,
\end{equation}
where $b_0$ is a constant which determines the length of the extra dimension.
In this model, the 4-dimensional Planck scale is given by
\begin{equation}
\label{4D-Planck-scale}
M_{\rm P}^2 = \frac{M^3}{k}[1-e^{-kb_0}],
\end{equation}
and for $\frac12kb_0\gtrsim1$ it is $k$ rather than $\frac12b_0$ that determines it.
RS \cite{RS1} argued that, due to the warp factor $e^{-kb_0y}$
which has different values at the hidden brane ($y=0$)
and at the visible brane ($y=\frac12$),
any mass parameter $m_0$ on the visible brane corresponds to
a physical mass $m=m_0e^{-\frac12kb_0}$ and
the moderate value $\frac12kb_0\sim37$ can produce
the huge ratio $M_P/M_W\sim10^{16}$.
Thus the gauge hierarchy problem is converted to
the problem related to geometry, fixing the size of the extra dimension.
Then it is an important question how precise the relation (\ref{RS-condition})
must be in order for the RS solution to work 
since the exact relation is assumed a priori 
to get (\ref{RS-metric}).

Now we try to answer this question by solving the Einstein equations
with the bulk and the brane cosmological constants,
but without the fine tuned condition (\ref{RS-condition}).
Boundary condition (\ref{BC}) suggests $n=a$, and we first
try a separable function $n=a=g(\tau)f(y)$. Then the (05) equation yields
$b=b(y)$. After separate coordinate transformations of $\tau$ and $y$,
we come to an ansatz
\begin{equation}
n=f(y), \quad a=g(\tau)f(y), \quad b=b_0,
\end{equation}
where $b_0$ is a constant.
Now subtracting the (ii) equation by the (00) equation,
we obtain $(\dot g/g)\dot{\vphantom{g}}=0$.
So we define $(\dot g/g)\equiv H_0={\rm constant}$.
Then the (55) equation gives
\begin{equation}
\left(\frac{f'}{b_0}\right)^2 = H_0^2 + k^2 f^2,
\end{equation}
and the (00) and (ii) equations just give a redundant equation.
For $\Lambda_b<0$, the solution to this equation consistent with the orbifold
symmetry is
\begin{equation}
f = \frac{H_0}{k}\sinh(-kb_0|y|+c_0).
\end{equation}
The boundary condition (\ref{BC}) imposes
\begin{equation}
\label{inflation-BC}
k_1 = k\coth(c_0), \quad
-k_2 = k\coth(-\frac12kb_0+c_0).
\end{equation}
Therefore, the solution is allowed when $k_1,k_2$ and $k$ satisfy
$k<k_1<-k_2$ and they are related to the length of the extra dimension
$L_5=\frac12b_0$ by
\begin{equation}
\label{inflation-condition}
L_5 = \frac{1}{2}b_0 =
\frac{1}{2k}\ln\left[\frac{-k_2-k}{k_1-k}\cdot\frac{k_1+k}{-k_2+k}\right].
\end{equation}
The metric of this solution is
\begin{eqnarray}
\label{inflation-metric}
ds^2 &=& \left(\frac{H_0}{k}\right)^2\sinh^2(-kb_0|y|+c_0)
\select{}{\nonumber\\&&\times}
\left[-d\tau^2+e^{2H_0\tau}\delta_{ij}dx^idx^j\right]
+b_0^2dy^2.
\end{eqnarray}
We arrive at the static limit by taking  $H_0\rightarrow0$ and
$c_0\rightarrow\infty$ while keeping the ratio
$\frac{H_0}{2k} e^{c_0}\rightarrow1$ fixed,
and the metric (\ref{inflation-metric}) becomes (\ref{RS-metric}).
For $H_0\neq0$, $H_0$ is not a physical quantity and can be set to k
which corresponds to the shift of the initial value of $\tau$.
Then by the coordinate transformation
$dy=d\tilde{y}/\sinh(kb_0|\tilde{y}|+\tilde{c}_0)$,
the metric becomes \cite{Nihei}
\begin{equation}
\label{inflation-metric1}
ds^2 = \frac{-d\tau^2+e^{2k\tau}\delta_{ij}dx^idx^j+b_0^2d{\tilde y}^2}
{\sinh^2(kb_0|\tilde{y}|+\tilde{c}_0)}.
\end{equation}

The metric (\ref{inflation-metric}) describes inflation of the spatial
dimensions with the length of extra dimension fixed.
At a given $y$, we can perform a 4-dimensional coordinate transformation
to make the 4-dimensional metric be in the form
$ds_{(4)}^2=-dt^2+e^{2H(y)t}\delta_{ij}dx^idx^j$.
Then we get the Hubble parameter
\begin{equation}
H(y) = k\,{\rm csch}(-kb_0|y|+c_0).
\end{equation}
Especially, at each boundary, we have
\begin{equation}
\label{hubble}
H(0) = \sqrt{k_1^2-k^2}, \quad
H(\frac12) = \sqrt{k_2^2-k^2},
\end{equation}
respectively.
We see that inflation occurs when the bulk and the brane cosmological
constants deviate from the relation (\ref{RS-condition}), which
can be easily seen from the boundary equation (\ref{Friedman-equation}).
The condition (\ref{inflation-condition}) means that
to keep the length of the extra dimension fixed while the spatial dimensions
inflate, we must fine tune the bulk and the brane cosmological constants,
in other words, we must put two branes a distance $L_5$ apart
for given $k_1$, $k_2$ and $k$.

For the more general case that the distance between two
branes is larger or smaller than $L_5$,
the solution can be obtained with the non-separable function
\begin{equation}
n(\tau,y)=a(\tau,y)=\frac{1}{\tau f(y)+g_0}, \quad
b(\tau,y)=k b_0\tau a(\tau,y),
\end{equation}
where $b_0$ and $g_0$ are constants.
The metric for this general case can be found to be
\begin{equation}
ds^2 = \frac{-d\tau^2+\delta_{ij}dx^idx^j+(k b_0\tau)^2dy^2}%
{\left[k \tau\sinh(kb_0|y|+c_0)+g_0\right]^2},
\end{equation}
where $b_0$ and $c_0$ are given by
\begin{eqnarray}
c_0 &=& \cosh^{-1}\left(\frac{k_1}{k}\right), \nonumber\\
kb_0 &=& 2\left[\cosh^{-1}\left(\frac{-k_2}{k}\right)
              -\cosh^{-1}\left(\frac{k_1}{k}\right)\right].
\end{eqnarray}
When $g_0=0$, this metric becomes (\ref{inflation-metric1}) by the
coordinate transformation  $\frac{1}{\tau}\,d\tau = -k\,d\tilde\tau$.
When $g_0>0$ ($g_0<0$), the distance between two branes is
smaller (larger) than $L_5$ and the above metric describes that
two branes come close and finally meet (move away from each other)
while the spatial dimensions are inflating. The boundary condition is not
affected by the presence of $g_0$, and $H(0)$ and $H(\frac12)$ are the
same as (\ref{hubble}), as seen in (\ref{Friedman-equation}).

For the special case $k_1=-k2>k$, $L_5$ becomes 0 and the above solutions
cannot cover. Instead, we can find a solution with a different ansatz
$n(\tau,y)=a(\tau,y)=b(\tau,y)$. The metric is given by
\begin{equation}
ds^2 = \frac{-d\tau^2+\delta_{ij}dx^idx^j+dy^2}%
{\left[-(k_1^2-k^2)^{1/2}\tau+ky+c_0\right]^2},
\end{equation}
where $c_0$ is a constant.
This metric describes inflation in both the spatial dimensions and the extra
dimension.

Let us consider the connection between the inflationary solutions and
the RS static solution.
The static limit that both $k_1$ and $-k_2$ approach $k$
in the inflationary solutions corresponds to the RS solution.
Suppose that the five-dimensional universe underwent inflation
in the early epoch and finally settles down to the static RS model.
Then the relation (\ref{RS-condition}) does not hold exactly
but approximately now.
This situation is most likely described by the static limit of the
inflationary solutions.
Then, the current observations on the Hubble constant restrict
the visible brane Hubble parameter $H(\frac{1}{2})$:
\begin{equation}
H(\frac{1}{2}) = \sqrt{k_2^2-k^2} \lesssim 10^{-60} M_{\rm P}
\end{equation}
where $k = {\cal O}(M) = {\cal O}(M_{\rm P})$ is assumed.
Therefore, the bulk and the visible brane cosmological constants must
cancel each other up to very high precision.
This is a five-dimensional version of
the well-known cosmological constant problem \cite{Kaloper} and
the RS condition (\ref{RS-condition}) is nothing but the condition for the
vanishing 4-dimensional effective cosmological constant at both branes.
We do not attempt to solve this notorious problem in this paper.
Instead, we concentrate on the gauge hierarchy problem within this context.

The key point of the RS solution to the gauge hierarchy problem is
the size of the extra dimension appearing in the exponential warp factor.
If the RS condition (\ref{RS-condition}) exactly holds,
it is not determined in this framework and remains as a flat direction 
in the moduli space.
For the bulk and the brane cosmological constants that
do not satisfy the RS condition, there is a critical value $L_5$,
for which the extra dimension size remains constant.
Our general solution shows that this is an unstable stationary configuration.
A slight deviation will make the extra dimension size shrink or grow.
This is a generic consequence of gravity theory,
and there is a way to overcome it by including extra dynamics
beyond simple gravity for the modulus $b$.
This is a five-dimensional version of the modulus stabilization problem.
Toward the solution of this problem,
an attempt using the bulk scalar field was made recently in \cite{GW2} and
an earlier attempt within the compactified heterotic M theory
without the bulk cosmological constant was done in \cite{choi}
using the membrane instanton effects and the racetrack mechanism.
We will not discuss it which is beyond the scope of this paper.

With this in mind,
let us look at the RS solution for the gauge hierarchy problem.
In the static limit $k_1,-k_2 \rightarrow k$ of the inflationary solutions
with the fixed extra dimension size, we can see from
(\ref{inflation-condition}) that the extra dimension size dose not have a
unique value but varies depending on how $k_1$ and $-k_2$ approach $k$.
This is just what is said by the modulus stabilization problem.
Near the static limit, the length of the extra dimension is expressed
in terms of the Hubble parameters of two branes by
\begin{equation}
k L_{\rm RS} = \frac12 k b_{\rm RS} \simeq
\ln\frac{H(\frac12)}{H(0)}.
\end{equation}
Since $H(\frac12)$ is very small, to keep $\frac12kb_{\rm RS}$ of order 1,
we also have to adjust the bulk and the hidden brane cosmological constants
to the same accuracy as we did for the bulk and visible brane cosmological
constants.
The RS solution for the gauge hierarchy problem demands
$\frac12kb_{\rm RS}\sim37$,
and  this number seems quite moderate at first sight.
However, in fact, it requires further fine tune of the hidden brane Hubble
parameter than the visible one:
\begin{equation}
H(0) = \sqrt{k_1^2-k^2} \simeq 10^{-16}H(\frac12) \lesssim 10^{-76}M_{\rm P}.
\end{equation}
Thus, to solve the gauge hierarchy problem in the context of the RS model,
the balance between the bulk and the hidden brane cosmological
constants should take place with $10^{16}$ times more accuracy
than that between the bulk and the visible cosmological constants.
At any rate, the RS model again converts the gauge hierarchy to the fine
tuning of the bulk and the brane cosmological constants.
It will be interesting to look at the problem 
in terms of four dimensional effective theory
and see whether there is a nice stabilization mechanism by which the above
argument can be avoided
using additional interactions of the $b$ modulus other than the gravity.


In summary, we found the general inflationary solutions for the slab of
5-dimensional AdS spacetime with two boundary 3-branes
and viewed the RS model as the static limit of those solutions.
Then both the cosmological constant problem and the gauge hierarchy problem
are recast into the fine tuning problem
of the bulk and the brane cosmological constants.
The cosmological constant problem appear as the fine tuning between the bulk
and the visible brane cosmological constants, and the gauge hierarchy problem
as the more severe fine tuning between the bulk
and the hidden brane cosmological constants.
The inclusion of matter at the brane boundaries might alter the above
conclusion, but we expect it does not much.
The real solution of those problems surely requires additional
ingredients to the 5-dimensional model considered in this paper,
but the implications of this simple model is so interesting that it deserves
further study.
The magic bullet for the cosmological constant problem and
the gauge hierarchy problem may be found in the correlation mechanism
of the bulk and brane cosmological constants,
which reminds us of the recent development in string theory, the holography.
We also leave the modulus stabilization problem which
is connected to the problem addressed in this letter for future work.

\acknowledgements

We would like to thank Jai-chan Hwang,
Youngjai Kiem, Yoonbai Kim for useful discussions
and especially Kiwoon Choi for his valuable comments.

\select{}{\end{multicols}}
\end{document}